\documentclass[a4paper,11pt,onecolumn]{article}
\usepackage{cite}
\usepackage[T1]{fontenc}
\usepackage{graphicx}
\usepackage{authblk}
\usepackage{amsmath}
\usepackage[margin=1in]{geometry}

\title{Tailored supercontinuum generation using genetic algorithm optimized Fourier domain pulse shaping}
\date{}

\author[1,2]{\small Mathilde Hary}
\author[1]{\small Lauri Salmela}
\author[1]{\small Piotr Ryczkowski}
\author[1]{\small Francesca Gallazzi}
\author[2]{\small John M. Dudley}
\author[1,*]{\small Go\"ery Genty}

\affil[1]{\footnotesize Photonics Laboratory, Tampere University, FI-33104 Tampere, Finland}
\affil[2]{\footnotesize Universit\'{e} de Franche-Comt\'{e}, Institut FEMTO-ST, CNRS UMR 6174, 25000 Besan\c{c}on, France}

\affil[*]{goery.genty@tuni.fi}
\begin{document}

\maketitle
\begin{abstract}
We report the generation of spectrally-tailored supercontinuum using Fourier-domain pulse shaping of femtosecond pulses injected into a highly nonlinear fiber controlled by a genetic algorithm.  User-selectable spectral enhancement is demonstrated over the 1550-2000~nm wavelength range, with the ability to both select a target central wavelength and a target bandwidth in the range 1--5~nm.  The spectral enhancement factor relative to unshaped input pulses is typically $\sim$5--20 in the range 1550--1800~nm and increases for longer wavelengths, exceeding a factor of 160 around 2000~nm.  We also demonstrate results where the genetic algorithm is applied to the enhancement of up to four wavelengths simultaneously.
\end{abstract}

A supercontinuum is a versatile light source that has revolutionized many applications such as imaging, spectroscopy, and sensing \cite{DudleyTaylor,Labruyère2012}.  The dynamics of supercontinuum generation are highly nonlinear and complex, especially in the anomalous dispersion regime where the resulting spectral features are associated with ejected soliton pulses, dispersive waves generation, and spectral interference \cite{TaylorSC}. Although it is now routine to generate supercontinuum spectra with very broad bandwidths, obtaining a desired spectral coverage with a large fraction of intensity in particular wavelength bands is more challenging, often requiring time-consuming trial and error experiments as well as computationally demanding simulations.

Machine learning is showing great promise in enabling ``smart'' control of light sources \cite{GG_LS_ML,NahriExtrem,SalmelaPred}, and techniques such as genetic algorithms (GA) and neural networks have been applied to actively control the complex dynamics and output characteristics of pulses from fiber lasers \cite{Kutz2014,Tianprateep,ZhangHighCO,Arteaga-Sierra,Boscolo,Vikramadaptative}, extra-cavity pulse compression \cite{Shen2017}, and controlled spectral broadening in planar waveguides induced by multipulse sequences \cite{Wetzel2018}.  In this paper, we apply a genetic algorithm to the systematic optimization of supercontinuum generation in a highly nonlinear fiber (HLNF), focussing in particular on enhancing the spectral intensity over narrow bandwidths of 1~nm and 5~nm, selected arbitrarily over the wavelength range 1550-2000~nm. 

\bigskip
Our approach is based on computer-controlled Fourier-domain spectral shaping to adjust the phase of a few 100's~fs pulses injected into a highly nonlinear fiber. Specifically, a genetic algorithm optimizes the spectral phase of the input pulses so as to maximise the spectral intensity at one or more desired output wavelengths.  For single target wavelengths, we quantify the spectral enhancement resulting from the optimization as a function of the desired central wavelength, and show that the technique works best for longer wavelengths exceeding $\sim$1850~nm with enhancement factors in the range 10--160. We also show that the algorithm can be adapted for multi-wavelength enhancement with the simultaneous optimization of the spectral density at three and four wavelengths. We also describe the evolution properties of the algorithm which shows rapid convergence to the desired target regime in $\sim$20 generations for single wavelength optimization, and after $\sim$50 generations for the multi-wavelength case.  These results provide a further demonstration of the power of machine learning techniques in harnessing complex dynamical processes for particular applications in photonics.

\begin{figure}[t]
\centering
\includegraphics[width=0.9\linewidth]{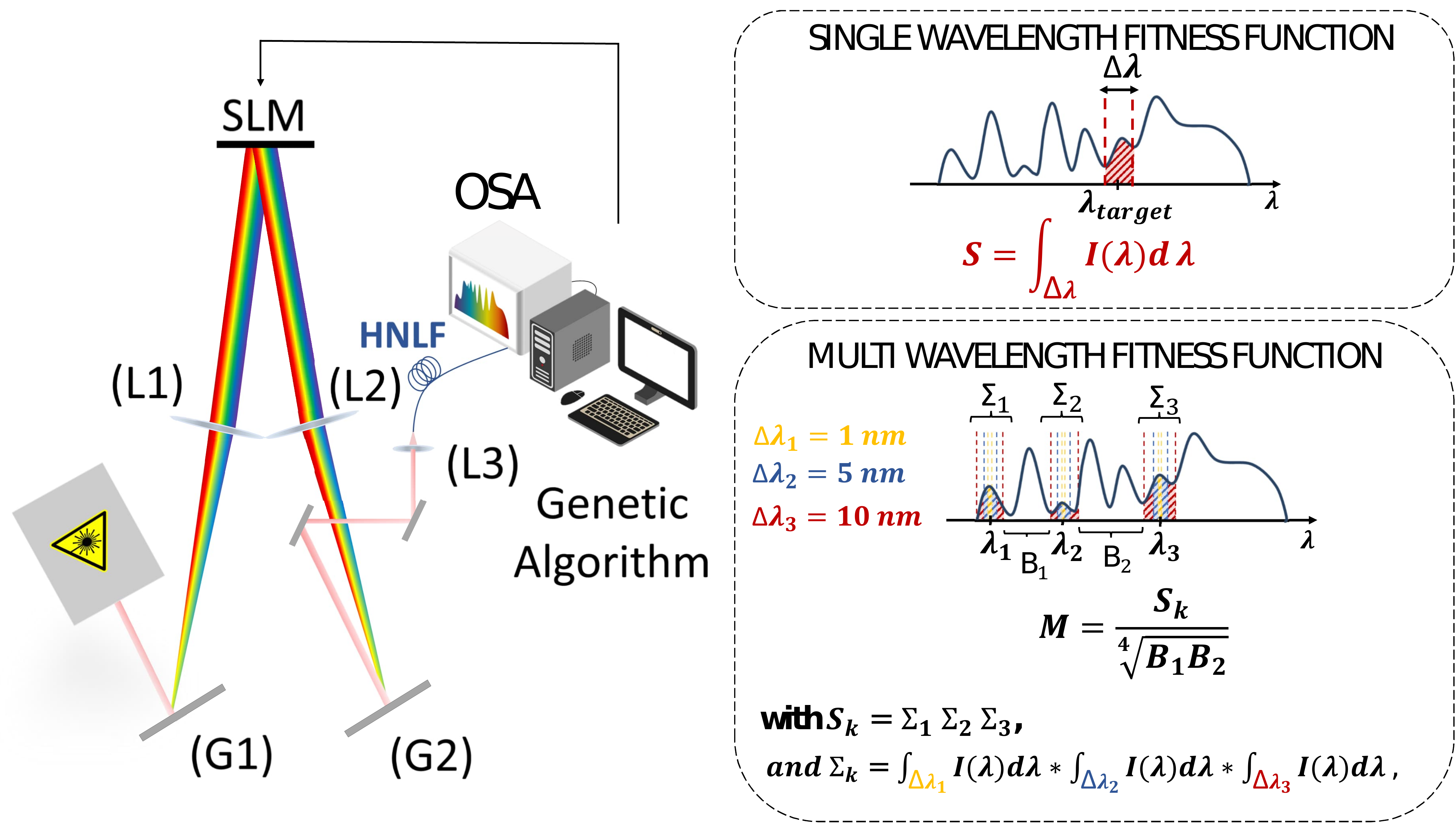}
\caption{Schematic of the pulse shaping apparatus and the feedback loop from the genetic algorithm to the spatial light modulator (SLM). The setup is composed of identical gratings (G1, G2, 600 lines/mm, $D = 1.46\,\mathrm{nm/mrad}$) and identical lenses (L1, L2, f = 30~cm). Light is collected after the pulse shaper and focused into a highly nonlinear fiber.  The fiber output is coupled into an optical spectrum analyzer (OSA) and the spectral intensity characteristics are optimized by a genetic algorithm which controls the spectral phase applied to the SLM. The fitness functions used for single and multi-wavelength optimization (see main text for a detailed description) are also highlighted in the figure.}
\label{fig1}
\end{figure}

The spectral phase imposed by the SLM is controlled by a genetic algorithm to enhance in target wavelength bands the spectral intensity of the supercontinuum generated in the highly nonlinear fiber. Note that evolutionary algorithms have been previously applied for the optimization of femtosecond and nanosecond laser pulses \cite{FarfanNNcomp,EfimovChirp,Tada,Baumert1997,SongFlat,Kashiwagi,Andral,Michaeli,Omenetto,Woodward,Pu_Intel,LapreGA,SalmelaIntegrator} and here, our approach is specifically adapted to the optimization of supercontinuum characteristics.

\bigskip

The algorithm evolves a population of 50 individuals, each individual specified by the five parameters of the system (genes). The SLM phase function (applied in the dispersed plane of the beam) is: $\phi_{\mathrm{SLM}}(\lambda) = c_1(\lambda-\lambda_0)^2 + c_2(\lambda-\lambda_0)^3+ c_3(\lambda-\lambda_0)^4 $, where the three coefficients (genes) $c_1,c_2,c_3$ quantify the spectral phase contributions up to fourth order (quartic), and the central frequency of the phase pattern $\lambda_0$ constitutes a fourth gene.  In addition, a fifth gene can be optimized to control the overall system throughput and hence the input power injected into the fiber.  This is achieved by applying a phase ramp to the SLM in the direction orthogonal to the grating dispersion  and tilting the beam, for example up. The angular tilt decreases the light coupling efficiency to the fiber \cite{FrumkerPhase}.

Starting from an input generation of randomly selected genes, we compute for each individual a fitness function \cite{Hauptbook}. In the case of single-wavelength optimization, this fitness function corresponds to the integrated spectral intensity around a specific target wavelength and over a specified bandwidth: $ S = \int I(\lambda) \mathrm{d}\lambda$. The individuals in the  subsequent generation are then selected using standard techniques of elitism (5\%), cross-over (9\%), and mutation (8\%). This process evolves through multiple generations until convergence is reached and there is no significant further improvement. The ``best'' individual is then retained as the algorithm output. Note that for speed of execution, the fitness function during the optimization steps for each individual is computed by scanning over a limited bandwidth of 30~nm around the target wavelength and the full spectrum is recorded only upon convergence. The implementation was done using the Global Optimization toolbox from MATLAB and optimization takes typically 25 minutes per wavelength optimization.

Figure 2 shows typical results targeting spectral optimization at 1900~nm and for target bandwidths of (a) $\Delta\lambda =1$~nm and (b) $\Delta\lambda =5$~nm. More specifically, Figs. 2(a) and (b) plot in blue the HNLF output supercontinuum spectra without optimization (i.e. using only the unshaped pulses) and with GA optimization (red and purple respectively). The insets show the corresponding spectra using a logarithmic scale in dB. The GA optimized results were obtained after 50 generations.  When comparing Fig 2(a) and (b), it is clear that, although the optimized spectra are different over the range 1500--1800~nm, the intensity characteristics around the target wavelength of 1900~nm are similar (with a spectrally broader feature in (b) as expected.)  

We repeated the optimization procedure above for target wavelengths in the full range 1550--2000~nm scanned in increments of 50~nm (and again for bandwidths of $\Delta\lambda =1$~nm and $\Delta\lambda =5$~nm.) It is convenient to quantify the optimization results here by defining a spectral intensity enhancement factor $\eta = S_f/S_i$ where $S_f$ and $S_i$ are the optimized and unoptimized spectral intensities integrated over the target bandwidths, respectively.  Figure 2(c) plots the enhancement factor as a function of target wavelength for bandwidths of $\Delta\lambda =1$~nm (red) and $\Delta\lambda =5$~nm (blue). The algorithm yields excellent performance with greater than $\times20$ enhancement above 1800~nm.  The results for both bandwidths show a similar trend. The improved performance at longer wavelengths is attributed to the fact that the particular spectral feature that is being optimized here is a distinct Raman-shifted soliton pulse which is separated from the more modulated spectral features closer to the pump.

\begin{figure}[!b]
\centering
\includegraphics[width=\linewidth]{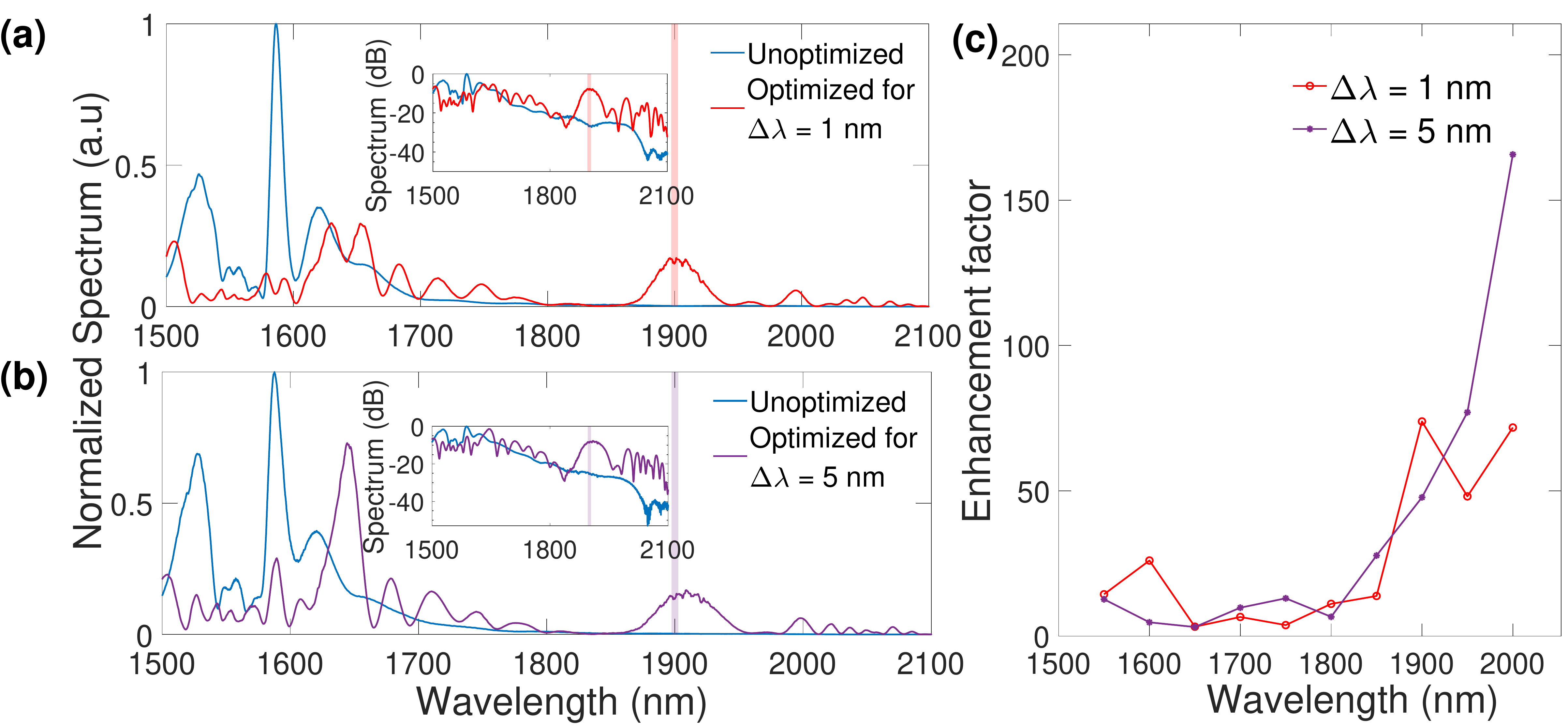}
\caption{(a) and (b) plot in blue the HNLF output spectra without optimization (i.e. generated by the unshaped pulses) and the HNLF output spectra with GA optimization for bandwidth of $\Delta\lambda = 1$ nm and $\Delta\lambda = 5$ nm in red and purple, respectively. The insets show the corresponding spectra in logarithmic scale.  (c) Spectral intensity enhancement factor vs. wavelength for single-wavelength optimization and for bandwidths of $\Delta\lambda = 1$ nm (red) and $\Delta\lambda = 5$ nm (purple). Spectra were normalized with respect to the maximum spectral intensity recorded over the measurement series.}
\label{fig:2}
\end{figure}

\bigskip
\begin{figure}[t]
\centering
\includegraphics[width=0.75\linewidth]{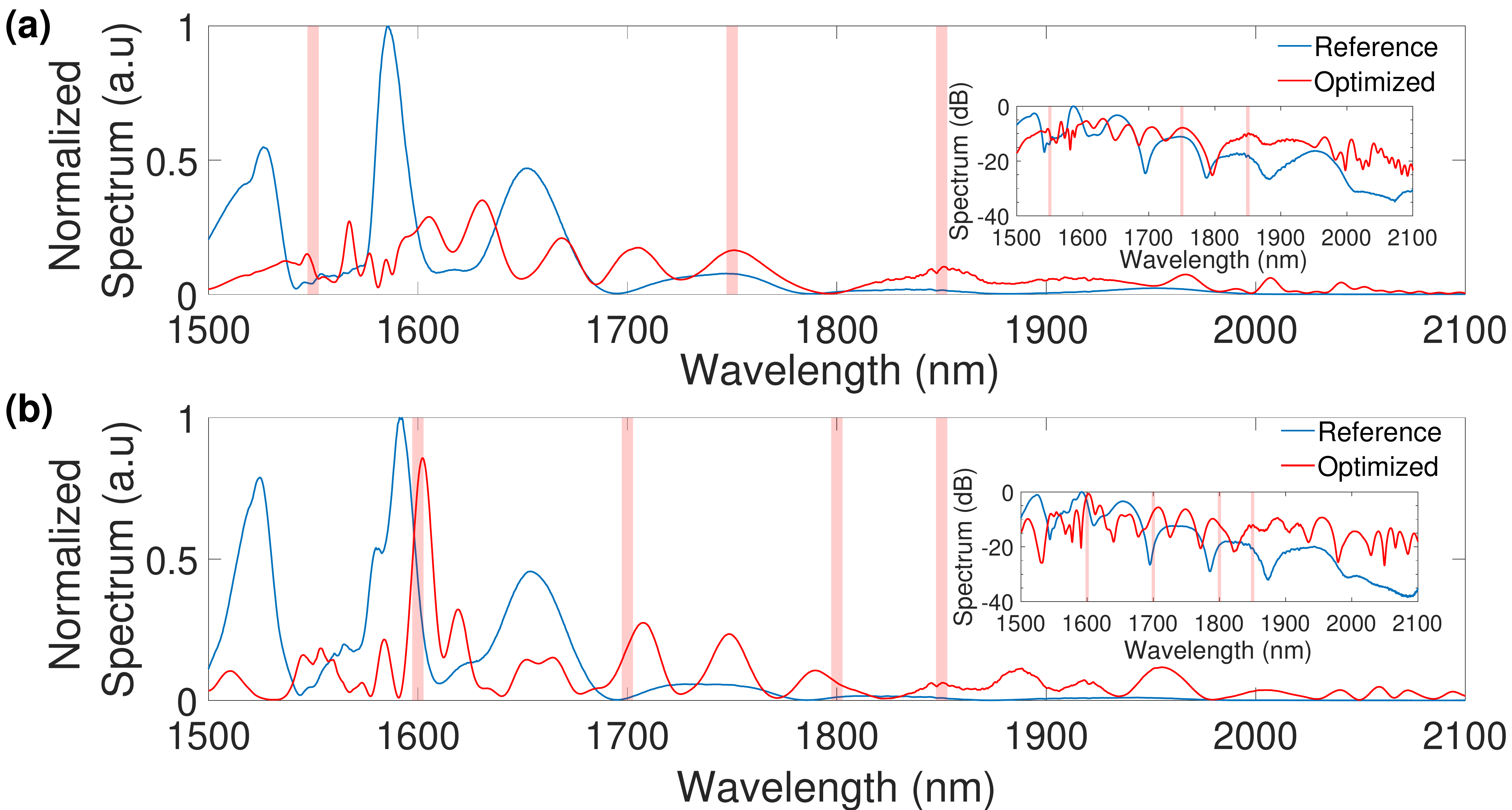}
\caption{Optimization of multiple wavelengths simultaneously, (a) at $1550$~ nm, $1750$~nm and $1850$~nm. (b) at $1600$~nm, $1700$~nm, $1800$~nm, $1850$~nm. Both figures show the spectrum before (blue) and after optimization by the GA (red). Note that spectrum before optimization can slightly change due to fluctuations in the pump laser power. The red rectangles indicate the position of the optimized wavelength bands.}
\label{fig:3}
\end{figure}

\begin{figure}[!b]
\centering
\includegraphics[width=0.7\linewidth]{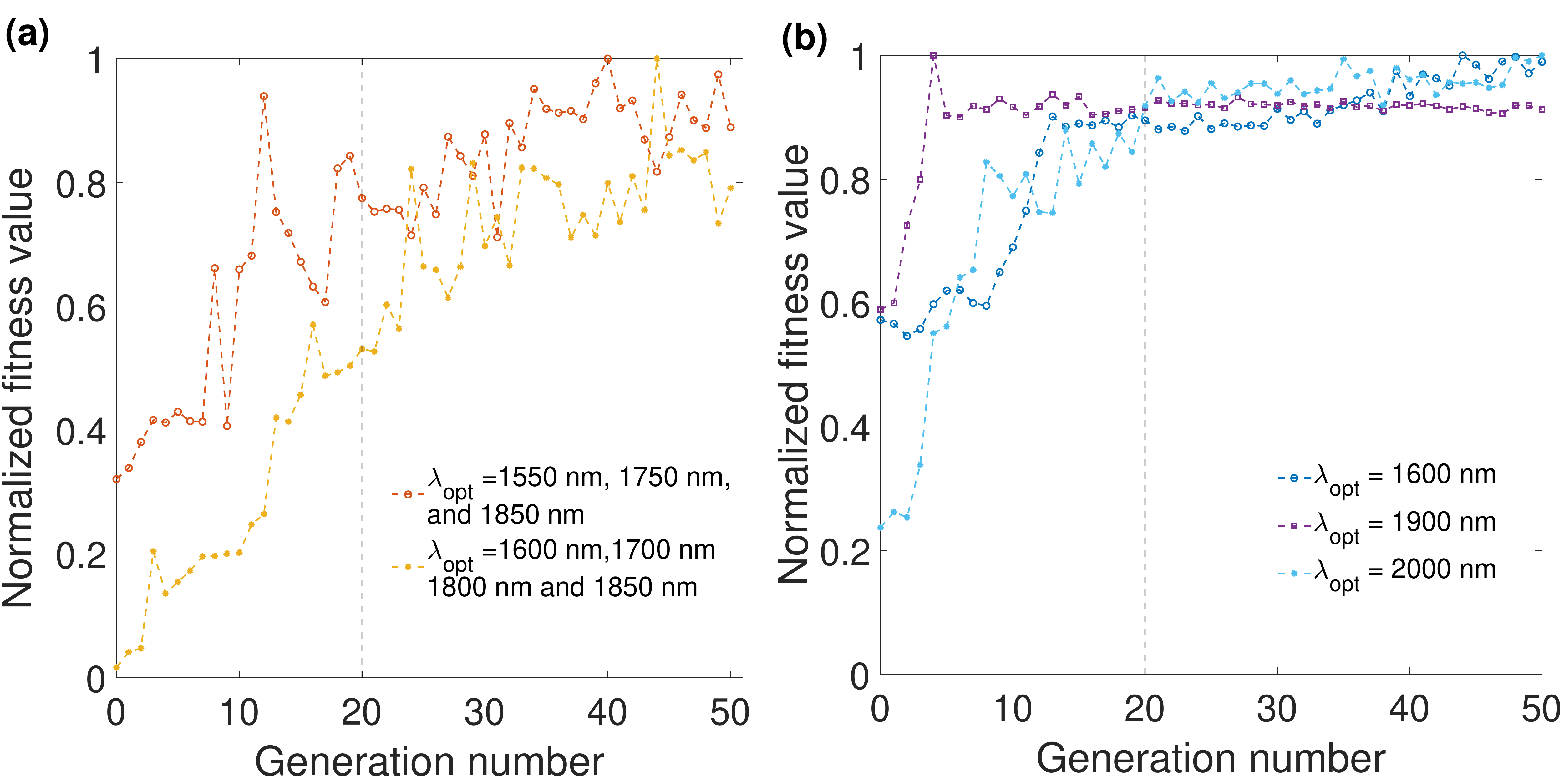}
\caption{(a) The red and yellow curves show the fitness function evolution for the simultaneous optimization of three and four different wavelengths, respectively. The vertical dashed lines mark the fitness value after 20 generations. (b) Optimization for single wavelength shown for three different wavelengths at 1600, 1900 and 2000 nm.}
\label{fig:4}
\end{figure}
For multi-wavelength optimization, we need to define a more complex fitness function that favours the enhancement of spectral characteristics at a number of different wavelengths simultaneously.  A challenge comes from the intrinsic property of the genetic algorithm: although we optimize multiple parameters, there is only one final fitness value. In particular, we need to be able to clearly force the algorithm to favor multiple peaks (rather than e.g. one peak with very large intensity), and we also need to be able to differentiate between a well-defined spectral peak with convex structure, and e.g. a concave dip in the spectrum with the same area over the target bandwidth. For the case where we seek spectral optimization at three wavelengths, the definition of the fitness function is illustrated in Fig.~1, with associated definition: $M = \Sigma_1 \Sigma_2 \Sigma_3 / (B_1 B_2)^{1/4}$. Here at each target wavelength $\lambda_k$, each $\Sigma_k = S_k(\Delta \lambda_1) S_k(\Delta \lambda_2) S_k(\Delta \lambda_3) $ in the numerator is the product of the integrated spectral intensity over increasing bandwidths $\Delta \lambda_1 = 1$~nm, $\Delta \lambda_2 = 5$~nm, $\Delta \lambda_3 = 10$~nm. By integrating over different bandwidths in this way, we favour the generation of convex peaks.  In the denominator, each $B_j$ quantifies the energy in the wavelength range between the target peaks as shown in Fig.~1. Maximising the fitness function then requires maximising the energy in all target peaks as well as minimising the energy between them.  The fourth root in the denominator was found empirically to yield the best convergence.  Figure~3(a) and Figure~3(b) show results of the optimization for three ($1550$~nm, $1750$~nm and $1850$~nm) and four wavelengths ($1600$~nm, $1700$~nm,$1800$~nm and $1850$~nm) simultaneously. Note that those wavelengths were selected arbitrarily for illustration purposes.

\bigskip

Typical results of the GA optimization procedure are shown in Fig.~4 for single and multiple wavelengths.  Specifically, over 50 generations, Fig.~4(a) plots the fitness value computed over the 50 individuals in the population for a three-wavelength optimization (red dots), as well as the fitness value for the best individuals in the case of four-wavelength optimization (yellow dots). For completeness, Fig.~4(b) shows the fitness value evolution over the generation for single wavelength optimization in three different cases ($\lambda = 1600$ nm, $\lambda = 1900$ nm and $\lambda = 2000$ nm).  The GA evolution shown in the figure reveals that the regime of convergence is reached rapidly when only a single wavelength is optimized.  Indeed, for these particular results, the algorithm appears to identify the optimal regime after only 20 generations, although the precise evolution for any particular experiment does depend on the initial genes that are selected randomly. Nonetheless, additional testing revealed that between 10-20 generations to enter into the optimal regime was typical.

The experiments reported here shows the strong dependence of the supercontinuum features on the input phase parameters applied before propagation into a highly nonlinear fiber. Automatic optimization using a genetic algorithm has been shown to effectively improve the spectral intensity for a given wavelength in a broad wavelength range and can be achieved in less than 30 min.  Our results here were demonstrated over 1550-2000~nm, a technical limit imposed by the optical spectrum analyzer.
Our results open up novel perspectives for light sources with on-demand spectra tailored to specific applications.

\bigskip

\noindent\textbf{{Funding.}} M.H, L.S., F.G. and G.G. acknowledge support from the Academy of Finland (Grants 318082, 333949, Flagship PREIN 320165).
J.M.D.acknowledges support from the French Investissements d'Avenir programme, project ISITE-BFC (contract ANR-15-IDEX-0003) and project EUR (ANR-17-EURE-0002).

\bigskip



\bibliographystyle{ieeetr}
\bibliography{Harybib}

\end{document}